\documentclass[10pt,aps,pre,twocolumn,superscriptaddress]{revtex4-2} 
\usepackage[english]{babel}
\usepackage[T1]{fontenc}
\usepackage{tikz}
\usepackage{amssymb,amsmath,amsfonts} 
\usepackage{textcomp} 
\usepackage{braket} 
\usepackage{nicematrix} 
\usepackage{graphicx,color}
\usepackage[utf8x]{inputenc}
\usepackage{bm}
\usepackage{mathbbol}
\usepackage{comment} 
\usepackage[normalem]{ulem}
\usepackage{graphicx}
\usepackage{listings}
\usepackage{stmaryrd}
\usepackage{amssymb}
\usepackage{amsmath}
\usepackage{gensymb}
\usepackage{bbold}
\usepackage{bm}
  
\usepackage[linkcolor = blue, citecolor = red, urlcolor = blue, colorlinks = true]{hyperref}

\usepackage{tabularx}

\begin{document}
 
\title{On the geometric algebras of the Ising model}
\author{N. Johnson}
\affiliation{School of Physics and Astronomy, The University of Edinburgh, James Clerk Maxwell Building, Peter Guthrie Tait Road, Edinburgh, EH9 3FD, United Kingdom}
\author{D. Marenduzzo}
\affiliation{School of Physics and Astronomy, The University of Edinburgh, James Clerk Maxwell Building, Peter Guthrie Tait Road, Edinburgh, EH9 3FD, United Kingdom}
\author{A. Morozov}
\affiliation{School of Physics and Astronomy, The University of Edinburgh, James Clerk Maxwell Building, Peter Guthrie Tait Road, Edinburgh, EH9 3FD, United Kingdom}
\author{E. Orlandini}
\affiliation{Department of Physics and Astronomy, University of Padova and  Sezione INFN, Padova, Via Marzolo 8, I-35131 Padova, Italy}
\author{G.~M. Vasil}
\affiliation{School of
Mathematics and Maxwell Institute for Mathematical Sciences, The University of Edinburgh, Edinburgh, EH9 3FD, United Kingdom}
 
\begin{abstract}
We revisit the classical transfer-matrix solution of the one- and two-dimensional Ising model from the perspective of Clifford and conformal geometric algebras. Building on Kaufman’s spinor formulation, we show that all elements entering the solution -- including the transfer matrix, its eigenvectors, and the quasiparticle excitations -- admit a natural and unified interpretation as elements of an appropriate conformal Clifford algebra. In particular, the transfer matrix can be viewed as a dilation generated by a conformal bivector, while its eigenvectors correspond to null combinations of Clifford generators, closely paralleling the emergence of Majorana fermionic degrees of freedom. In the two-dimensional case, the standard eigenvalue equation for the row-to-row transfer matrix is reinterpreted as a dispersion relation for quasiparticle excitations, exposing the connection between the Ising model and a theory of free Majorana fermions. While all the explicit exact results recovered are well known, this geometric reformulation provides a unified algebraic framework which is compact and  physically interpretable. Specifically, this clarifies the role of scale transformations, fermionic modes, and duality in the Ising model. We believe this approach offers a useful pedagogical complement to more conventional fermionic, Grassmann, or field-theoretic treatments.
\end{abstract}

\maketitle

\section*{Introduction}

The Ising model~\cite{mccoy1973} occupies a central position in statistical physics and lattice field theory as a paradigm for %one of the earliest and most paradigmatic 
exactly solvable models. Since Onsager’s original computation of the two-dimensional partition function using transfer-matrix techniques~\cite{onsager1944}, the model has served as a testing ground for analytical methods ranging from spinor calculus to conformal field theory and integrable systems \cite{kaufman1949,schultz1964,itzykson1982,cardy1996,mussardo2010}. Despite its apparent simplicity, the Ising model continues to provide insight into universality, critical phenomena, and the emergence of fermionic degrees of freedom and dual disorder variables~\cite{fradkin1978} in classical statistical systems.

%The Ising model~\cite{kaufman1949,onsager1944} is arguably the prototype of exactly solved models in statistical physics and lattice field theory~\cite{wolff2020}. While its solution was achieved by Onsager~\cite{onsager1944}, who computed the partition function by the transfer matrix approach, the technicality of the method renders reproducing the solution an analytical tour-de-force. An alternative solution was proposed shortly after by Kaufman, who, by using spinor calculus, was able to find a simpler way to get to the same solution~\cite{kaufman1949}. This approach at the same time exposes an intriguing link between the classical statistical mechanics model quantum-like quasiparticles, as the  excitations from the Ising model ground state may be viewed as fermionic or bosonic fields~\cite{schultz1964,kadanoff1971,jordan1928}.

A major simplification of Onsager’s solution was achieved shortly after by Kaufman, who reformulated the transfer-matrix problem using spinor analysis and Clifford algebra techniques~\cite{kaufman1949}. This approach revealed an underlying fermionic structure and paved the way for later developments, including the Jordan–Wigner transformation~\cite{jordan1928}, Grassmann integral formulations \cite{samuel1980}, and the interpretation of the two-dimensional Ising model as a theory of free Majorana fermions~\cite{schultz1964,itzykson1982,wolff2020}. These connections also underlie the scale invariance of the model at criticality and its description in terms of conformal field theory~\cite{cardy1996,chelkak2012}.

In parallel, Clifford and geometric algebras have long been recognised as powerful tools for encoding spinors, fermions, and geometric transformations into a unified algebraic language~\cite{hestenes1967,hestenes1984,lundholm2009}. In particular, conformal geometric algebra provides a natural representation of rotations, boosts, and dilations as exponentials of bivectors. While Clifford algebras already play a role in Kaufman’s original formulation, their geometric interpretation -- especially in a conformal setting -- has not been systematically exploited in the context of the Ising model.

%The spinor solution in~\cite{kaufman1949} is also important in mathematical physics because it provides a nice application of Clifford algebras, or more generally geometric algebras, to a specific physical system model. In this work, we review this Clifford algebra approach, and expand it to provide a more direct interpretation of {\it all} objects in the model, from the transfer matrix to the eigenvectors and the quasiparticle excitations, in terms of elements of the Clifford algebras. In particular, we show that the 2-dimensional conformal geometric algebra, Cl(1,1), is ideally suited to describe the system, which is intuitively appealing as the Ising model is scale-invariance, and can be described by conformal field theory, via the free-fermion action. As we shown, this framework provides a simple and natural way to represent the transfer matrix formalism and diagonalize it, leading to a geometric interpretation of the quasiparticle excitations in the system, which contain Majorana fermions and a dilaton.

The purpose of this work is to revisit the classical solution of the Ising model in one and two dimensions by using conformal geometric algebra. Our aim is not to derive new results but rather to provide a coherent geometric reinterpretation of the standard transfer-matrix formalism, in which all objects appearing in the solution -- transfer matrices, eigenvectors, and quasiparticle excitations -- are represented explicitly as elements of a suitable Clifford algebra, which we shall show to be a two-dimensional conformal geometric algebra (or a direct product of many copies of it). In this framework, for instance in 1D, the transfer matrix naturally appears as a dilation generated by a conformal bivector, while its eigenvectors correspond to null combinations of Clifford generators. This viewpoint offers a transparent algebraic interpretation of the emergence of Majorana fermionic modes and clarifies the role of scale transformations and duality in the Ising model.

We begin by applying this approach to the one-dimensional Ising model, where the structure is particularly transparent and serves as a pedagogical introduction. We then extend the construction to the two-dimensional square lattice, following Kaufman’s spinor method, and show that the eigenvalue problem for the transfer matrix can be reinterpreted as a dispersion relation for quasiparticle excitations. This makes explicit the close analogy between the Ising model and a free fermionic system, an analogy that becomes exact at criticality. Although closely related observations are well known in fermionic and field-theoretic formulations, the geometric algebra perspective highlights these features in a compact and intuitive manner. 

We believe this reformulation provides an instructive example of the power of geometric algebras in finding deep links between different areas of theoretical physics~\cite{hestenes1967}. We hope that this work may also serve as a stepping  
%We hope that this reformulation provides a useful pedagogical bridge between classical statistical mechanics, fermionic representations, and geometric algebra methods,  
stone toward the analysis of more complex lattice models with richer algebraic structure.

%Moving on to the 2D Ising model on the square lattice, we demonstrate that the transfer matrix for a row of $N$ spins can be written as an element of the direct product of $N$ copies of Cl(1,1). The eigenvalues and eigenvectors of the transfer matrix are then obtained through an analysis of cyclic matrices as in~\cite{kaufman1949}. We show that the eigenvalue equation is equivalent to a dispersion relation for the quasiparticle excitations, which, as expected, likens the Ising model to a quantum condensed matter system with free Majorana fermions. While none of the results obtained is new, the interpretation is original and we believe provides an instructive example of the power of geometric algebras in finding deep links between different areas of theoretical physics~\cite{hestenes1967}.

\section{1D Ising model}

Let us begin with the 1D Ising model. This is useful for practicing the framework and for seeing how Pauli matrices -- and hence Clifford algebras -- naturally arise in the solution. It also provides a surprisingly simple way to visualise Majorana modes and quasiparticles algebraically, in terms of elements of the conformal Clifford algebra Cl(1,1). 

\subsection{Transfer matrix formalism}

The transfer matrix ${\mathbf T}$ of the 1D Ising model can be written as follows,
\begin{eqnarray}\label{TM1}
    {\mathbf T} & = &  
    \begin{pmatrix}
        e^K & e^{-K}\\
        e^{-K} & e ^{K}
    \end{pmatrix} 
    = e^{K} \mathbb{1}+e^{-K} \sigma_x,
\end{eqnarray}
where $\mathbb{1}$ is the $2\times 2$ identity matrix and $\sigma_{x}$ is the first Pauli matrix. Then for uniform lattice with $n$ nodes and periodic boundary conditions the partition function $Z(K) = \text{tr}({\mathbf T}^{n})$.

The purpose of the paper is to highlight the power of abstract algebraic methods for solving long tedious calculations. The details of the particular Pauli matrix representation are unimportant for our purposes, all that is needed is 
\begin{align}
    \sigma_{x}^{2} \, = \, \mathbb{1},
\end{align}
which is sufficient to derive 
\begin{align}
    e^{K^{*}  \sigma_{x}} \, = \, \cosh(K^{*})\, \mathbb{1} + \sinh(K^{*})\, \sigma_{x}, 
\end{align}
for a (yet undetermined) ``dual'' coupling constant, $K^{*}$ with the goal of imposing ${\mathbf T} = T_{0}\, e^{K^{*} \sigma_{x}}$ for another undetermined factor, $T_{0}$.  It is straightforward to find the dual parameters by the elementary properties $\sinh(K^{*}) / \cosh(K^{*}) = \tanh(K^{*})$ and $\cosh(K^{*})^{2} - \sinh(K^{*})^{2} = 1$. The $K^{*}$ parameter is ``dual'' in the sense of the involutive relationship,
\begin{align}
    \tanh(K^{*}) \, = \,   e^{-2K} \quad \iff \quad  \tanh(K) \, &= \,   e^{-2K^{*}}.
\end{align}
The proportionality factor is $T_{0} = \sqrt{2 \sinh(2K)}$.

\subsection{Clifford algebra representation}

In this Section we link the 1D Ising model to Clifford algebras. As mentioned in the Introduction, our goal is to provide, wherever possible an explicit representations for all the objects in the theory (transfer matrix, eigenvectors, etc.) as elements of a suitable algebra, as this exposes their geometric and physical meanings. 

The spin order parameter is here a simple scalar in $\mathbb{Z}_2$, so Cl(0,1) would suffice to describe it.  
However, to describe scale transformations (dilations), we make the algebra conformal -- and equivalent to Cl(1,1) -- by introducing two other generators, $e_+$, and $e_-$, which square to $+\mathbb{1}$ and $-\mathbb{1}$ respectively, 
\begin{eqnarray}
   & &  e_+^2=\mathbb{1}, \quad e_-^2=-\mathbb{1} \\ \nonumber 
   & &  e_{+}e_{-} = -e_{-}e_{+}\equiv e_{+-}. 
\end{eqnarray}
Note that $e_{+-}^{2} = \mathbb{1}$, making $\text{Cl}(1,1) \simeq \text{Cl}(2,0)$.
Wherever required, an explicit representation can be found in terms of $2\times 2$ real matrices as follows,
\begin{equation}
    e_{+}=\begin{pmatrix} 1 & 0 \\ 0 & -1 \end{pmatrix}, \; e_{-}=\begin{pmatrix} 0 & 1 \\ -1 & 0\end{pmatrix}, \; e_{+-}=\begin{pmatrix} 0 & 1 \\ 1 & 0  \end{pmatrix}.
\end{equation}
Note that $e_{+-}=\sigma_x$ and $e_{+} = \sigma_{z}$ in this representation. 

Clifford algebras also admit a natural geometric algebra structure using the inner (dot) and outer (wedge) products, of any two elements, 
\begin{align}
    e \cdot f \, = \, \tfrac{1}{2}( ef + fe), \qquad e \wedge f \, = \, \tfrac{1}{2}( ef - fe).  
\end{align}
Starting from $e_{\pm}$, we define two null vectors $e_0$ and $e_{\infty}$ as follows,
\begin{align}
    e_0 \, = \,  \tfrac{1}{2} (e_{+}+e_{-}), \qquad 
    e_{\infty} \, = \, e_{-} - e_{+}.
\end{align}
Null here means that these $e_{0}^{2} = e_{\infty}^{2} = 0$. Also, 
\begin{equation}
e_{0}\cdot e_{\infty}= - \mathbb{1}, \qquad e_{\infty}\wedge e_0 \equiv e_{\infty 0}= 
e_{+-}.
\end{equation}
Note that $e_\infty e_0 = -\mathbb{1} + e_{\infty 0}$.

We now need to find how to represent dilations in the Ising conformal algebra %, which for us is 
Cl(1,1). As shown in Appendix A, for a generic Clifford algebra Cl(n,0), the
bivector $e_{\infty 0}$ generates dilations or contractions in Cl(n+1,1). For the present case, $n=0$, we can therefore represent dilations $\Delta$ as elements of a group generated by $e_{\infty 0}$, as follows
\begin{equation}
    \Delta=e^{\alpha e_{\infty 0}}=\cosh(\alpha)+\sinh(\alpha)\,e_{\infty 0},
\end{equation}
which again relies on the fact that $e_{\infty 0}^{2} = \sigma_{x}^{2} = \mathbb{1}$. It is a simple matter to show that combinations of dilations are dilations and that $\Delta^{-1}=e^{-\alpha e_{\infty 0}}$. Dilations (or contractions) are the only relevant geometric transformations in the Ising model, and we shall see that they feature multiple times at various places, in both the emerging physics and the analytical solution of the model. 

\subsection{Diagonalisation of the transfer matrix}

In the interpretation of the Ising model solution, we only need dilations from the conformal geometric algebra, as these are the transformations that are equivalent to the transfer matrix operation. The eigenvectors of the transfer matrix can be identified with quasiparticles, which are dilated or contracted by the transfer matrix. 

For the transfer matrix in the form ${\mathbf T} = T_{0} e^{K^{*} \sigma_{x}}$ the eigenvectors of ${\mathbf T}$ coincide with those of $\sigma_{x}$. The Pauli matrices all have eigenvalues $\pm 1$, hance the transfer matrix has eigenvalues $T_{0} e^{\pm K^{*}}$. The eigenvectors of $\sigma_{x}$ are straightforward to find using a direct calculation, 
\begin{eqnarray}\label{v+-}
    {\mathbf v_{\pm}} = \tfrac{1}{\sqrt{2}} \left(\, \ket{+} \pm \ket{-} \,\right) \, = \, \tfrac{1}{\sqrt{2}}\begin{pmatrix} 1 \\ \pm 1
\end{pmatrix},
\end{eqnarray}
with the standard basis 
$\ket{+}=\begin{pmatrix} 1 \\ 0
\end{pmatrix}$ and 
$\ket{-}=\begin{pmatrix} 0 \\ 1 
\end{pmatrix}$. 
%Additionally, ${\mathbf v_+}$ can be qualitatively associated with a pair of nearest neighbours with equal spins ($++$ or $--$), whose weight gets inflated under the action of the transfer matrix, while ${\mathbf v_-}$ can be associated with a pair of nearest neighbours with unequal spins ($+-$ or $-+$), whose weight gets shrunk by $\tilde{\mathbf{T}}$.
 
Now we show how the Clifford algebra naturally furnishes not only a representation for operators matrices (\textit{i.e.}, $\sigma_{x}$) but also for the vectors, ${\mathbf v_{\pm}}$. A main philosophy of the Clifford algebra approach is that \textit{everything} is an element of the same algebra.  

The transfer matrix normalised ${\tilde{\mathbf T}} = {\mathbf T} / T_{0}$ is readily recognised as a dilation in Cl(1,1), as
\begin{equation}
    {\tilde {\mathbf T}} = \cosh(K^*)\,\mathbb{1}+\sinh(K^*)\,e_{+-}.
\end{equation} 
The eigenvalues are still $e^{\pm K^{*}}$. However, note that based \textit{purely on the abstract algebra of $e_{\pm}$} (without using the matrix representation), we have 
\begin{eqnarray}
    {\tilde {\mathbf T}} \, \xi_{\pm}
    & = &  e^{\pm K^{*}} \xi_{\pm}, \quad \text{for} \quad  \xi_{\pm} = \tfrac{1}{\sqrt{2}}(e_{+} \mp e_{-}),
\end{eqnarray} 
which follows from purely formal algebraic rules and does not require any specific matrix representation.

The ${\bf v}_{\pm}$ vectors are best viewed as spinors -- for instance, this can be recognised by noting they are $2\times 1$ column vectors in Eq.~(\ref{v+-}). 
However, at this point, $\xi_{\pm}$ are full matrices that only act like eigenvectors. We can multiply them from the right by a projector to obtain
\begin{equation}
    \psi_{\pm}  =   \xi_{\pm}P_{+} = \tfrac{1}{\sqrt{2}}\begin{pmatrix}
        1 & 0 \\
        \pm 1 & 0
    \end{pmatrix},
\end{equation}
where $P_+=\frac{\mathbb{1}+e_+}{2}$ (so that $P_+^2=P_+$).

The overall concept is that, in Clifford algebras, spinors are minimal left ideals~\cite{hestenes1984}, which means that for any element in the algebra, $\text{span}(e \psi_{\pm})$ is contained in $\text{span}(\psi_{\pm})$ (which is what happens naturally with matrices acting on a basis). In Cl(1,1), minimal left ideals can be represented by $2\times 2$ matrices where only the first column is non-zero. 

\begin{figure}
\begin{center}
\includegraphics[width=\columnwidth]{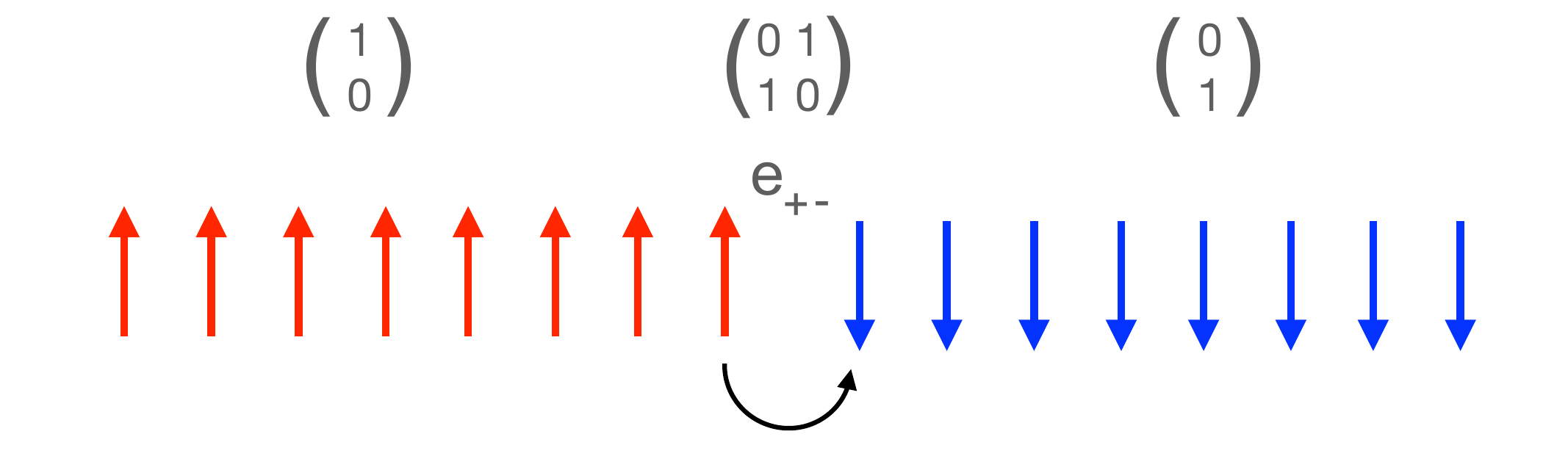}
\caption{Sketch of a domain wall in the 1D Ising model, which can be associated with an $e_{+-}$ bivector acting on the edge which separates the two domains. This is because $e_{+-}=\begin{pmatrix}
 0 & 1 \\ 1 & 0 \end{pmatrix}$ acts as a spin flip operator.}
\label{fig:domainwall}
\end{center}
\end{figure}

It is also sometimes customary to represent spinors as even elements of the algebra [which are a combination of the scalar and bivector in Cl(1,1)] -- these are the so-called ``Hestenes spinors''~\cite{hestenes1967,hestenes1984}. The Hestenes spinors corresponding to the eigenvectors can be found by multiplying $\xi_{\pm}$ (or in general any odd element in a Clifford algebra representation of an eigenvector) by $e_{+}$. This is because $e_{+} P_{+} = P_{+}$: hence, right multiplication by $e_+$ of an odd element of the algebra gives an even element without affecting the action of right multiplication by $P_+$, so that the resulting minimal left ideal obtained after projection by $P_+$ is the same. The Hestenes spinors, $\chi_{\pm}$, for the 1D Ising transfer matrix are therefore:
\begin{eqnarray}
    \chi_{\pm} = \frac{1}{\sqrt{2}}\left(\mathbb{1} \pm e_{+-}\right), \qquad \psi_{\pm} = \chi_{\pm} P_{+}
\end{eqnarray}
which shows they are a combination of unity and the dilation bivector, consistent with them being even elements. It can be verified that, once more, ${\tilde {\mathbf T}} \chi_{\pm} = e^{\pm K^{*}} \chi_{\pm}$, so this alternative representation of the eigenvectors also makes sense in the Clifford algebra. The element $e_{+-}$ is a spin flip operator, which is relevant here as it makes apparent that $\chi_{\pm}$ is related to a quasiparticle, here a domain wall singularity, which is the defect in the Ising model. For this analogy to hold we have to think of $e_{+-}$ applied to an edge between nearest neighbours than a single site, as only in that case can it be viewed as creating a domain wall by switching $\ket{+}\leftrightarrow \ket{-}$ (Fig.~\ref{fig:domainwall}).

Now it is useful to interpret (or reformulate) all the Clifford algebra elements we have encountered up to now in terms of objects defined in and familiar from quantum condensed matter theories. First, we have seen that the transfer matrix ${\mathbf{T}}=e^{K^*e_{+-}}$ can be identified with a dilation in the conformal geometric algebra.
Second, the elements $\gamma_1=e_{+}$ and $\gamma_2=ie_{-}$ can be viewed as Majorana zero modes in quantum condensed matter language, as they square to $\mathbb{1}$ and can be linearly composed to get eigenvectors in the Ising model -- equivalently, ``fermions'' in quantum condensed matter, a reasonable analogy as we have seen that eigenvectors are best viewed as spinors. Indeed, the eigenvectors $\xi_{\pm}$ can be written as combinations of these elements as $\xi_{\pm}=\frac{1}{\sqrt{2}}(\gamma_1\pm i\gamma_2)$: they are therefore analogous to fermionic destruction (or creation) operators in second quantisation jargon. These elements are also nilpotent, as they square to $0$.  [This is due to the mixed metric signature, and can be readily checked explicitly -- e.g., $\xi_{+}^2=\tfrac{1}{2}\left(e_{+}^2+e_{-}^2+e_{+}e_{-}+e_{-}e_{+}\right)=\tfrac{1}{2}(\mathbb{1}-\mathbb{1}+0)=0$.] If we naturally define a fermion creation operator as $c=\xi_{+}$ and an annihilation operator as $c^{\dagger}=\xi_{-}$, then the combination $cc^{\dagger}=\mathbb{1}+e_{+-}$ is the Hestenes spinor corresponding to the larger eigenvalue $\lambda_+$. 
Quite remarkably, finally, the bivector of the theory, $e_{+-}$, admits two complementary interpretations: 
as a dilation operator, qualitatively reminiscent of a dilaton of quantum conformal field theory, and as a spin flip operator which creates domain walls, the defects (or quasiparticles) of the Ising model.

Notably, all elements in the problem have explicit representations in the conformal Clifford algebra. One may quite reasonably point out that this approach is clearly not needed if all we need to do is solve the Ising model in 1D, as this can be done in a line without introducing any of this formalism. However, the Clifford framework remains useful for exposing the fundamental links this simple statistical physics model has to both quantum condensed matter and particle physics. Additionally, it proves instructive to the generalisation to 2D, which we consider next, as it will lead to a simpler solution and interpretation of the solution of the 2D Ising model with respect to the original exact solution~\cite{onsager1944}.
%and the dilation operator $e_{+-}$ is qualitatively reminiscent of the dilaton (a boson). %The bosonic nature of this quasiparticle is also apparent when considering its $n$-th power,
%    \begin{equation}
%        \left[\mathbb{1}+e_{+-}\right]^n=2^{n-1}\left[\mathbb{1}+e_{+-}\right]
%    \end{equation}
%\end{enumerate}

\section{2D Ising model}

\subsection{Transfer matrix formalism}

The partition function for the Ising model on the 2D square lattice, without a magnetic field, is given by
\begin{equation} 
    Z = \sum_{\rm configurations} e^{-E_c/(k_BT)},
\end{equation} 
where $E_c$ denotes the energy of a configuration, or microstate, and the partition function involves summation over all possible configurations or microstates of the system. 

It is useful to introduce row-configurations, labelled by $\{\nu_i\}_i$, where $i=1,\ldots,n$ is the row index. A row-configuration $\nu_i$ is given by $\{S_{i,1},\ldots, S_{i,n}\}$, with $S_{i,j}$ the value of the $j-$th spin in the row under consideration, $i$. We associate $\nu_i$ with a $2^n$ vector $\ket{\nu_i}$, whose entries are the values of the spin $S_{i,j}$, $\ket{\nu_i} = (S_{i,1},\ldots,S_{i,n})^T$ [and $\bra{\nu_i}=(S_{i,1},\ldots,S_{i,n})$]. Representing each spin also with bra-ket notation, this is equivalent to writing $\ket{\nu_i}$ as a tensor product,
\begin{equation}
    \ket{\nu_i} = \ket{S_{i,1}}\otimes \ket{S_{i,2}}\otimes \cdots \otimes \ket{S_{i,n}}
\end{equation}

The contribution to $E_c$ given by interactions within the $i$-th row is
\begin{equation}
    E_1(\nu_i) = -K\sum_{j=1}^{n}{S_{i,j}S_{i,j+1}},
\end{equation}
where we use periodic boundary conditions so that we can identify $S_{i,n+1}=S_{i,1}$. We can also define a $2^n\times 2^n$ matrix ${\bf T_1}$ that is diagonal, with diagonal elements given by:
\begin{equation}
    e^{-E_1(\nu_i)}\equiv \bra{\nu_i}{\mathbf T_1}\ket{\nu_i}.    
\end{equation}

The contribution to $E_c$ coming from interactions between the $i$-th row and the $(i+1)$-th row is instead given by:
\begin{equation}
    E_2(\nu_i,\nu_{i+1}) = -K\sum_{j=1}^{n}{S_{i,j}S_{i+1,j}}.
\end{equation}
Similarly to the case of $E_1$, we can define a $2^n\times 2^n$ matrix ${\bf T_2}$ whose non-zero matrix elements are given by:
\begin{equation}
    e^{-E_2(\nu_i,\nu_{i+1})}\equiv \bra{\nu_i}{\mathbf T_2}\ket{\nu_{i+1}}.    
\end{equation}

It is also useful to write ${\mathbf T_1}$ and ${\mathbf T_2}$ as a tensor product of $n$ $2\times 2$ matrices. For ${\mathbf T_1}$, we obtain~\cite{kaufman1949}:
\begin{eqnarray} 
    {\mathbf T_1} & \propto & e^{K^* \sum_{r=1}^n e_+^{(r)} e_{+}^{(r+1)}} \\ \nonumber
    e_{+}^{(r)} & = & \mathbb{1}\otimes \cdots \otimes \mathbb{1}  \otimes e_{+}\otimes \mathbb{1}\otimes \cdots \otimes \mathbb{1}, %\\ \nonumber
    %& = & (2\sinh{2K})e^{K^* \sum_{r=1}^n e_+^{(r)} e_{+}^{(r+1)}}
\end{eqnarray}
where we have dropped a multiplicative factor (dependent on $K$ but not on the Clifford generators) in the formula for ${\mathbf T_1}$. In the definition of $e_{+}^{(r)}$ the $r$-th factor in the direct product equals $e_{+}$ (and the rest of the factors equal $\mathbb{1}$). For ${\mathbf T_2}$, we get
\begin{eqnarray}
    {\mathbf T_2} & = & e^{K \sum_{r=1}^n e_{+-}^{(r)}} \\ \nonumber
    e_{+-}^{(r)} & = & \mathbb{1}\otimes \cdots \otimes \mathbb{1}  \otimes e_{+-}\otimes \mathbb{1}\otimes \cdots \otimes \mathbb{1},
\end{eqnarray}
where again in $e_{+-}^{(r)}$ the non-identity factor, $e_{+-}$, appears in the $r$-th term.

The partition function can be written as 
\begin{eqnarray}
    Z & = & \sum_{\ket{\nu_1}, \cdots, \ket{\nu_n}} \bra{\nu_1}{\mathbf T_2}\ket{\nu_2} 
    \bra{\nu_2}{\mathbf T_1}\ket{\nu_2} \cdots \\ \nonumber
    & & \cdots  
    %\bra{\nu_{n-1}}{\mathbf T_2}\ket{\nu_n} 
    \bra{\nu_n}{\mathbf T_1}\ket{\nu_n} 
    \bra{\nu_{n}}{\mathbf T_2}\ket{\nu_1} 
    \bra{\nu_1}{\mathbf T_1}\ket{\nu_1}\\ \nonumber
    & = & \sum_{\ket{\nu_1}} \bra{\nu_1} \left({\mathbf T_2 \mathbf T_1}\right)^n \ket{\nu_1}
    ={\rm Tr}\left[\left({\mathbf T_2}{\mathbf T_1}\right)^n\right]\, .
\end{eqnarray}
Therefore, our goal is to find ${\rm Tr}\left[\left({\mathbf T_2}{\mathbf T_1}\right)^n\right]$.

\subsection{Clifford algebra representation}

We define a $2n$-dimensional Clifford algebra Cl(n,n), by taking the direct product of $n$ copies of the Cl(1,1) algebra generated by $e_{+}$ and $e_{-}$, as follows,
\begin{eqnarray}
    \Gamma_{2r-1} & = & e_{+-}\otimes e_{+-} \otimes \cdots \otimes e_{+} \otimes \mathbb{1} \otimes \cdots \otimes \mathbb{1} %\equiv {\mathbf P_r}
    \\ \nonumber
    \Gamma_{2r} & = &  e_{+-}\otimes e_{+-} \otimes \cdots \otimes e_{-} \otimes \mathbb{1} \otimes \cdots \otimes \mathbb{1} %\equiv {\mathbf Q_r}
\end{eqnarray}
where $1\le r\le n$, and where the factors of $e_{+}$ or $e_{-}$ appear in the $r$-th factor in the direct product.

The %${\mathbf P_r}$ 
odd $\Gamma_{2r-1}$ generators square to $+\mathbb{1}$, whereas the %${\mathbf Q_r}$ 
even $\Gamma_{2r}$ generators square to $-\mathbb{1}$, 
\begin{equation}
    %{\mathbf P_r}^2=+1 \, , \qquad {\mathbf Q_r}^2=-1\, . 
    \Gamma_{2r-1}^2=+\mathbb{1} \, , \qquad  {\Gamma_{2r}^2}=-\mathbb{1}\, . 
\end{equation}
It can also be seen that any two different generators anticommute, as required by the definition of a Clifford algebra, 
\begin{equation}
    \Gamma_k \Gamma_l + \Gamma_l \Gamma_k =0 \qquad {\rm if} \, k\ne l.
\end{equation}

By using bivectors made up by one odd generator %of type ${\mathbf P}$ 
and one even generator, %of type ${\mathbf Q}$,   
we can construct dilation operators, such as 
\begin{equation}
    {\mathbf S}=e^{K\Gamma_{2k-1}\Gamma_{2l}}=\cosh{K} + \Gamma_{2k-1}\Gamma_{2l}\sinh{K}\, .
\end{equation}
There are $n^2$ distinct choices for such dilations. 

A transformation which is important for finding the partition function of the 2D Ising model is a combination of dilations made up by $n$ pairs of separate bivectors (each constructed with one odd and one even generator), %type ${\mathbf P}$ and one of type ${\mathbf Q}$), 
which constitute $n$ commuting dilations. There are $n!$ choices of these. We consider the case in which each of the dilations acts on the same $r$, so are performed by the bivector ${\Gamma_{2r-1}}{\Gamma_{2r}}$, or equivalently $e_{+-}^{(r)}$. For one dilation of this type, the following equality holds
\begin{equation}
    e^{K{\Gamma_{2r-1}}{\Gamma_{2r}}} = %& = & \left[\cosh{K} \mathbb{1}\otimes \cdots \mathbb{1} + \sinh{K} \mathbb{1}\otimes \mathbb{1} \otimes \cdots \otimes e_{+-}\otimes \mathbb{1} \otimes \cdots \otimes \mathbb{1}\right] \\ \nonumber
    %& = & \mathbb{1} \otimes \mathbb{1} \otimes \cdots \otimes \left[\cosh{K}+\sinh{K} e_{+-}\right]\otimes \mathbb{1}\otimes \cdots \otimes \mathbb{1} \\ \nonumber 
    %& = & 
    \mathbb{1} \otimes \mathbb{1} \otimes \cdots \otimes e^{Ke_{+-}}\otimes \mathbb{1}\otimes \cdots \otimes \mathbb{1}
\end{equation}
Therefore, the product of dilations, ${\mathbf D_n}$, can be written as 
\begin{eqnarray}
  {\mathbf D_n} & = & \prod_{i=1}^n e^{K_i {\Gamma_{2i-1}}{\Gamma_{2i}} } \\ \nonumber 
  %& = &  \left[\cosh{K_1} +\sinh{K_1}e_{+-} \right]\otimes \cdots \otimes \left[\cosh{K_n}+\sinh{K_n} e_{+-}\right] \\ \nonumber 
  & = & e^{K_1e_{+-}}\otimes \cdots \otimes e^{K_ne_{+-}}  
\end{eqnarray}

Now we can find the eigenvalues by diagonalising ${\mathbf D_n}$ term by term in the direct product to obtain
\begin{eqnarray}
    {\mathbf \Lambda_n} = 
    \begin{pmatrix}
        e^{K_1} & 0 \\
        0 & e^{-K_1} 
    \end{pmatrix}
    \otimes \cdots \otimes
    \begin{pmatrix}
        e^{K_n} & 0 \\
        0 & e^{-K_n}
    \end{pmatrix}\, .
\end{eqnarray}
The $2^n$ eigenvalues of ${\mathbf D_n}$ (equivalently ${\mathbf \Lambda_n}$) are therefore given by: 
\begin{equation}\label{eigDn}
    \lambda_j = e^{\pm K_1\pm \cdots \pm K_n}
\end{equation}
for $j=1,\ldots, 2^n$: each eigenvalue corresponds to one independent choice of either $+$ or $-$ for each of $i=1,\ldots,n$. We will use this result to compute the eigenvalues of ${\mathbf T_2}{\mathbf T_1}$ in the following Section. 

It is important to realise that the matrix ${\mathbf D_n}$ can also be written as a bilinear form in Clifford algebra as $\tfrac{1}{2}{\mathbf \Gamma^\dagger}{\mathbf{\tilde{D}_n}}{\mathbf \Gamma}$, with
\begin{eqnarray}
    {\mathbf \Gamma}^\dagger & = & (\Gamma_1,-\Gamma_2,\ldots,\Gamma_{2n-1},-\Gamma_{2n}) \\ \nonumber
    {\mathbf \Gamma} & = & (\Gamma_1,\Gamma_2,\ldots,\Gamma_{2n-1},\Gamma_{2n})^T \\ \nonumber
    {\mathbf{\tilde{D}_n}} & = & \begin{pmatrix}
        {\cosh K_1} & {\sinh K_1} & \ldots  & 0 & 0\\
        {\sinh K_1} & {\cosh K_1} & \ldots  & 0 & 0\\
        0 & 0 & \ldots &  0 & 0\\
        0 & 0 &  \ldots &  0 & 0\\
        \ldots & \ldots & \ldots & \ldots & \ldots \\
         0 & 0 &  \ldots & \cosh K_n & {\sinh K_n} \\
         0 & 0 &  \ldots & \sinh K_n & {\cosh K_n}
    \end{pmatrix} \, .
\end{eqnarray}
The eigenvalues of ${\mathbf{\tilde{D}_n}}$ are then given by $e^{\pm K_1},\cdots,e^{\pm K_n}$, which translate to the eigenvalues in Eq.~(\ref{eigDn}) for ${\mathbf D_n}$. Note that the negative signs in ${\mathbf \Gamma^{\dagger}}$ are due to the metric of the Cl(n,n) algebra, which is not positive definite. It is often simpler in practice to diagonalise the matrix in ``Clifford algebra space'' -- i.e., to diagonalise the $2n\times 2n$ matrix ${\mathbf{\tilde{D}}_n}$ -- instead of the $2^n\times 2^n$ matrix ${\mathbf D}_n$), and indeed we shall use this shortcut in the 2D Ising model calculation below.

\subsection{Diagonalisation of the transfer matrix} %{Eigenvalue calculation}

To find the trace of ${\mathbf T_2} {\mathbf T_1}$, we need some properties of cyclic matrices. This is because both ${\mathbf T_1}$ and ${\mathbf T_2}$ %, as well as ${\mathbf T_2}^{1/2}{\mathbf T_1}{\mathbf T_2}^{1/2}$ 
are cyclic matrices. We follow~\cite{kaufman1949} closely in this Section. 

Let us consider the following generic cyclic matrix ${\mathbf C}$,
\begin{equation}
    {\mathbf C} = 
    \begin{pmatrix}
        a_1 & a_2 & a_3 & \ldots & a_{n-1} & a_n\\
        a_2 & a_3 & a_4 & \ldots & a_{n} & a_1\\
        a_3 & a_4 & a_5 & \ldots & a_1 & a_2 \\
        \ldots & \ldots & \ldots & \ldots & \ldots & \ldots \\
         a_n & a_1 & a_2 & \ldots & a_{n-2} & a_{n-1},
    \end{pmatrix} 
\end{equation}
where $\{a_i\}_{i=1,\ldots,n}$ are scalars. 
The eigenvectors of ${\mathbf C}$ are given by the following formula, which is essentially in the form of a discrete Fourier transform:
\begin{equation}
    {\mathbf v} = 
        \begin{pmatrix}
        \epsilon^r\\
        \epsilon^{2r}\\
        \epsilon^{3r} \\
        \ldots  \\
        \epsilon^{(n-1)r}\\
        1
    \end{pmatrix},
    \qquad \epsilon=e^{i \frac{2\pi}{n}}, \qquad 1\le r \le n \, .
    \end{equation}
This can be checked, for instance, by direct substitution.
The corresponding eigenvalues are given by:
\begin{equation}
    \lambda_r = a_1+a_2\epsilon^r+a_3\epsilon^{2r}+\ldots+a_{n-1} \epsilon{(n-2)r}+a_n \epsilon^{(n-1)r}.
\end{equation}

Let us now consider the generalisation where ${\mathbf C}$ is given by matrices, as follows,
\begin{equation}
    {\mathbf C} = 
    \begin{pmatrix}
        {\mathbf a_1} & {\mathbf a_2} & {\mathbf a_3} & \ldots & {\mathbf a_{n-1}} & {\mathbf a_n}\\
        {\mathbf a_2} & {\mathbf a_3} & {\mathbf a_4} & \ldots & {\mathbf a_{n}} & {\mathbf a_1}\\
        {\mathbf a_3} & {\mathbf a_4} & {\mathbf a_5} & \ldots & {\mathbf a_1} & {\mathbf a_2} \\
        \ldots & \ldots & \ldots & \ldots & \ldots & \ldots \\
         {\mathbf a_n} & {\mathbf a_1} & {\mathbf a_2} & \ldots & {\mathbf a_{n-2}} & {\mathbf a_{n-1}},
    \end{pmatrix} 
\end{equation}
where $\{{\mathbf a_i}\}_{i=1,\ldots,n}$ are $k\times k$ matrices. 

The eigenvectors in this case are given by:
\begin{equation}
    {\mathbf v} = 
        \begin{pmatrix}
        \epsilon^r {\mathbf w_r}\\
        \epsilon^{2r} {\mathbf w_r}\\
        \epsilon^{3r} {\mathbf w_r}\\
        \ldots  \\
        \epsilon^{(n-1)r} {\mathbf w_r}\\
        {\mathbf w_r}
    \end{pmatrix},
    \qquad \epsilon=e^{i \frac{2\pi}{n}}, \qquad 1\le r \le n \, .
\end{equation}
where ${\mathbf w_r}$ is an eigenvector of the $k\times k$ matrix:
\begin{equation}
    {\mathbf A_r}={\mathbf a_1}+\epsilon^{r} {\mathbf a_2}+\epsilon^{2r} {\mathbf a_3}+\ldots \epsilon^{2(n-1)r}{\mathbf a_n}.
\end{equation}
The eigenvalues $\lambda_r$ of ${\mathbf A_r}$ are also the eigenvalues of ${\mathbf C}$.

In our case, we need to compute the  eigenvalues of the $2n \times 2n$ matrix representation (in Clifford space) of the transformation induced by ${\mathbf T_2}^{1/2}{\mathbf T_1}{\mathbf T_2}^{1/2}$. The matrix ${\mathbf T_2}$ is associated with the following dilation in $2n\times 2n$ space:
\begin{equation}
    \begin{pmatrix}
        {\cosh k^*} & {\sinh k^*} & \ldots  & 0 & 0\\
        {\sinh k^*} & {\cosh k^*} & \ldots  & 0 & 0\\
        0 & 0 & \ldots &  0 & 0\\
        0 & 0 &  \ldots &  0 & 0\\
        \ldots & \ldots & \ldots & \ldots & \ldots \\
         0 & 0 &  \ldots & \cosh k^* & {\sinh k^*} \\
         0 & 0 &  \ldots & \sinh k^* & {\cosh k^*}
    \end{pmatrix} \, ,
\end{equation}
whereas ${\mathbf T_1}$ is associated with the dilation:
\begin{equation}
    \begin{pmatrix}
        {\cosh k} & 0 & 0 & \ldots  & 0 & 0 & \sinh k\\
       0 &  {\cosh k} & {\sinh k}  & \ldots & 0 &  0 & 0\\
        0 & {\sinh k} & {\cosh k} & \ldots & 0 & 0 & 0\\
        \ldots & \ldots & \ldots & \ldots & \ldots & \ldots & \ldots \\
         0 & 0 & 0 & \ldots & \cosh k & {\sinh k} & 0 \\
         0 & 0 & 0 & \ldots & \sinh k & {\cosh k} & 0\\
         {\sinh k} & 0 & 0 & \ldots & 0 &  0 & {\cosh k}
    \end{pmatrix} \, .
\end{equation}

We now need to find ${\rm Tr}({\mathbf T_2}\mathbf{T_1})$. By the cyclic property of the trace, or equivalently by its symmetric property that ${\rm Tr}(AB)={\rm Tr}(BA)$, we can instead consider the matrix ${\mathbf W}={\mathbf T_2}^{1/2}\mathbf{T_1}{\mathbf T_2}^{1/2}$, as its trace is the same as that of ${\mathbf T_2}\mathbf{T_1}$. 

The matrix ${\mathbf W}$ can be found to be
\begin{equation}
    \begin{pmatrix}
        {\mathbf A} & {\mathbf B_1} & {\mathbf 0} & \ldots & {\mathbf 0} & {\mathbf B_2}\\
        {\mathbf B_2} & {\mathbf A} & {\mathbf B_1} & \ldots & {\mathbf 0} & {\mathbf 0}\\
        {\mathbf 0} & {\mathbf B_2} & {\mathbf A} & \ldots & {\mathbf 0} & {\mathbf 0} \\
        \ldots & \ldots & \ldots & \ldots & \ldots & \ldots \\
         {\mathbf B_1} & {\mathbf 0} & {\mathbf 0} & \ldots & {\mathbf B_2} & {\mathbf A},
    \end{pmatrix} 
\end{equation}
with
\begin{eqnarray}
   {\mathbf A} & = &
   \begin{pmatrix}
   \cosh 2K^* \cosh 2K & \sinh 2K^* \cosh 2K \\
   \sinh 2K^* \cosh 2K & \cosh 2K^* \cosh 2K
   \end{pmatrix} \, , \\ \nonumber
   {\mathbf B_1} & = & 
   \begin{pmatrix}
   \frac{1}{2}\sinh 2K^* \sinh 2K & \sinh^2 K^* \sinh 2K \\
   \cosh^2 K^* \sinh 2K & \frac{1}{2}\sinh 2K^* \sinh 2K
   \end{pmatrix} \, , \\ \nonumber
   {\mathbf B_2} & = & 
   \begin{pmatrix}
   \frac{1}{2}\sinh 2K^* \sinh 2K & \cosh^2 K^* \sinh 2K \\
   \sinh^2 K^* \sinh 2K & \frac{1}{2}\sinh 2K^* \sinh 2K
   \end{pmatrix} \, . 
%   = {\mathbf B_1}^T\, .
\end{eqnarray}

Note that $\mathbf{B}_2 = \mathbf{B}_1^T$. The eigenvalues of ${\mathbf W}$ are also the eigenvalues of the $2\times 2$ matrix 
\begin{equation}
    {\mathbf M} = {\mathbf A}+\epsilon^r {\mathbf B_1} + \epsilon^{-r} {\mathbf B_2} \, ,
\end{equation}
which are given by $\lambda_r=e^{\pm \alpha_r}$ (they need to multiply to $1$ as ${\rm det} (\mathbf M)=1$), with
\begin{eqnarray}\label{dispersion}
    \cosh \alpha_r  & = & \frac{1}{2}{\rm Tr}({\mathbf M}) \\
    \nonumber 
    %& = &  \cosh 2K^*\cosh 2K +\sinh 2K^* \sinh 2K \cos{\left(\frac{2\pi r}{n}\right)} \\ \nonumber
    & = & \cosh 2K^*\cosh 2K +\cos{\left(\frac{2\pi r}{n}\right)} \, ,
\end{eqnarray}
where in the last line we have used the duality relation $\sinh 2K^* \sinh 2K=1$.
Eq.~(\ref{dispersion}) is related to the cosine rule in hyperbolic geometry (with an angle $\pi-\frac{2\pi r}{n}$ between the segments of size $2K$ and $2K^*$), which is relevant as dilations are formally equivalent to boosts, which are composed by summing vectors in hyperbolic space.

As ${\mathbf M}$ is a $2\times 2$ matrix, its ($r$-dependent) Clifford algebra eigenvectors can be given in terms of $e_{+}$ and $e_{-}$ as follows:
\begin{eqnarray}
    {\mathbf w}_r^{(\pm)} & = &  \frac{1}{\sqrt{2}} \left(e_{+}\pm \frac{z_r}{|z_r|} e_{-}\right) \\ \nonumber
    z_r & = & \left(\sinh{2K^*}\cosh{2K}+\cosh{2K^*}\sinh{2K}\cos{\frac{2\pi r}{n}}\right) \\ \nonumber
    & & +i\left(\sinh{2K^*}\sin{\frac{2\pi r}{n}}\right)
\end{eqnarray}
Using the generators of Cl(n,n) as a suitable basis, the eigenvectors of the transfer matrix for a given $r$ are therefore
\begin{eqnarray}
    {\mathbf v}_r^{(1)} & = & \sum_{k=1}^n \frac{e^{i2\pi rk/n}}{\sqrt{2}}\left[{\Gamma}_{2k-1}+\frac{z_r}{|z_r|}\Gamma_{2k}\right] \\ \nonumber
    {\mathbf v}_r^{(2)} & = & \sum_{k=1}^n \frac{e^{i2\pi rk/n}}{\sqrt{2}}\left[{\Gamma}_{2k-1}-\frac{z_r}{|z_r|}\Gamma_{2k}\right].
\end{eqnarray}
These eigenvectors correspond to the two values of the eigenvalues, $\lambda_r=e^{\pm \alpha_r}$, for a given $r$.

\subsection{Dispersion relation and physical interpretation}

\begin{figure*}
\begin{center} 
\includegraphics[width=2.0\columnwidth]{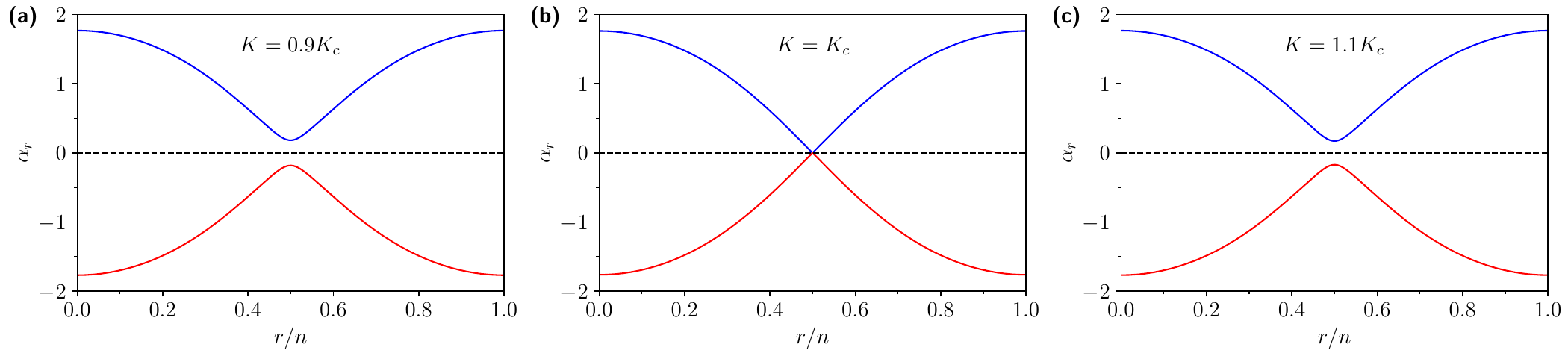}
\caption{Dispersion relation showing the angle $\alpha_r=\log(\lambda_r)$ (logarithm of the eigenvalues of ${\mathbf M}$ and of ${\mathbf W}$) versus $r/n$ (normalised wavenumber). (a) Low $K$ (high temperature) phase at $K=0.4$: there is a gap in the system. (b) Critical behaviour at $K=K_c\simeq 0.4407$ (defined by $K_c^*=K_c$): the gap closes, and a zero-energy mode (low-energy quasi-particle excitation) appears around $r=n/2$. (c) High $K$ (low temperature) phase: the gap has reopened.}
\label{fig:dispersion}
\end{center}
\end{figure*}

Eq.~(\ref{dispersion}) can also be thought of as a dispersion relation, as $r$ is a wavenumber, and the eigenvectors are discretised Fourier modes. Plotting the two values of $\alpha_r$ as a function of $r$ then gives the analogue of two bands in a quantum material (Fig.~\ref{fig:dispersion}). The gap between the two eigenvalues provides a measure of the inverse correlation length, or the inverse size of the quasi-particle excitations, which can be thought of as dilations corresponding to introducing an island of the wrong phase; for instance an island of negative spins in a sea of positive spins in the high $K$ (low temperature) phase. The mass of the quasi-particles becomes zero at criticality, where the system is scale invariant, and hence pattern dilations cost no energy. We can therefore think about the quasi-particles corresponding to the excitations around $r=n/2$ as dilations, or dilaton-like particles, whose mass is finite in the high $K$ and low $K$ phase, and vanishes at the critical point $K=K_c$. 

We now analyse the dispersion relation, Eq.~(\ref{dispersion}), close to $q_r\equiv 2\pi r/n=\pi$, where the gap closes at $K=K_c$. Let us first define an effective mass $m$ via
\begin{equation}
    m^2 = 2\left(\cosh{2K^*}\cosh{2K}-1\right).
\end{equation}
It can be seen that $m=0$ at $K=K_c$. The dispersion relation can be recast in terms of $\alpha_r$ as follows, 
\begin{equation}
    \alpha_r  =  {\rm acosh}\left[1+\frac{m^2}{2}+\cos(q_r)\right].
\end{equation}
For $q_r=-\pi+\delta q_r$ and small $q_r$ (close to the gap), we can use the fact that ${\rm acosh}(1+x)\sim \sqrt{2x}$ for $x\to 0$ to obtain
\begin{equation}
\alpha_r \sim \sqrt{m^2+\delta q_r^2},
\end{equation}
which can be recognised as a relativistic dispersion relation for the energy of a particle with mass $m$ and momentum $\delta q_r$ (with speed of light $c=1$). For $K\ne K_c$, the dispersion is quadratic close to the gap, whereas it becomes linear for $K=K_c$. When $r=n/2$ ($q_r=\pi$), the eigenvectors also simplify, they become real and are given by a combination of ${\Gamma}_{2k-1}+{\Gamma}_{2k}$, and of ${\Gamma}_{2k-1}-{\Gamma}_{2k}$, corresponding to ``light-like'' null vectors. 

The interpretation of the algebraic quantities close to the gap are therefore very similar to the one discussed for the 1D Ising model.
%\begin{enumerate} 
%   \item 
First, the elements ${\Gamma}_{2r-1}$ and $i\Gamma_{2r}$ can be viewed as $n$ pairs for Majorana zero modes, each corresponding to ``half'' a fermion.
Second, the eigenvectors $v_{+}$ and $v_{-}$ can be viewed as fermionic quasiparticles, as each is made up of a combination of $2$ Majorana modes. They are also null vectors (as we are looking at the system close to the gap). These quasiparticles are highly non-local; they are physically strings of domain walls and are mathematically equivalent to the fermions obtained by the Jordan-Wigner transformation~\cite{jordan1928}. 
Third, the elements $\mathbb{1}+e_{+-}^{(r)}$ can be viewed as bosonic quasiparticles, which cause local dilations, and can be qualitatively thought of as a dilaton-like particle. Such ``dilatons'' are therefore local islands of spins, which can be formed by combining two fermionic strings with the same endpoints. %The bosonic nature of this quasiparticle is also apparent when considering its $n$-th power,
%    \begin{equation}
%        \left[\mathbb{1}+e_{+-}\right]^n=2^{n-1}\left[\mathbb{1}+e_{+-}\right]
%    \end{equation}
%\end{enumerate}

\section{Conclusions}

In summary, we have used %an application of Clifford algebra, in particular 
geometric conformal algebras to study the Ising model in one and two dimensions.
Besides reviewing the classical solution of~\cite{kaufman1949} reinterpreting the spinors as conformal bivectors, our approach provides a geometric interpretation of the transfer matrix and its eigenvectors in terms of algebraic elements, allowing for a geometric understanding of the quasiparticle excitations in the system. In particular, we find that the transfer matrix can be written as a dilation in the conformal algebra, while the eigenvectors can be expressed in terms of the generators and are naturally interpreted as Majorana operators or zero modes. More importantly, the single bivector in the system has a dual interpretation. On the one hand, it generates dilations, similar to the dilaton of quantum conformal field theories. On the other hand, it %a string of bivectors 
can naturally be associated with the quasiparticle in the theory, which is a domain wall. Quite remarkably, all the algebraic structure is already present in the 1D version of the model, showing that this structure is general and intrinsically determined by the basic geometry and symmetry of the system.

In 2D, we show that it is possible to reinterpret the formula for the eigenvalues as a dispersion relation which likens this simple classical spin system to quantum condensed matter, where the physics is determined by the presence or absence of a gap in the quasiparticle dispersion relation. Here the gap closes, as expected, at the transition, and the low-energy excitations recover the free Majorana fermions which are well known to feature in this model. 

While our treatment recovers well-known results and does not yield anything fundamentally new in terms of exact results, it shows that conformal Clifford algebras provide a valuable tool to interpret geometrically and physically all quantities that enter the statistical physics of the problem, allowing us to draw an intriguing parallel to quantum condensed matter and topological band theory. We expect that more complex spin models, such as the Potts model~\cite{fradkin1980} or the Blume-Emery-Griffith model~\cite{blume1971}, will lead to a richer geometric structure, potentially opening up new avenues for exploring complex systems and their emergent phenomena. 

\begin{acknowledgments}
For the purpose of open access, the authors have applied a Creative Commons Attribution (CC BY) licence to any Author Accepted Manuscript version arising from this submission.
\end{acknowledgments}

\appendix
\renewcommand{\theequation}{A\arabic{equation}} % Custom format: A1, A2
\setcounter{equation}{0} % Reset counter

\subsection*{Appendix A: Dilations as bivectors in Cl(n+1,1)}

For a generic Clifford algebra Cl(n,0) with $n$ generators and dimension $2^n$, we can consider the following map between the vectors in Cl(n,0), labelled with lowercase letters, say $x$, and the vectors in the conformal extension Cl(n+1,1), labelled with capital letters, say $X$:
\begin{equation}\label{compactification}
    x \mapsto X(x)=x+e_0-\frac{x^2}{2}e_{\infty}.
\end{equation}
While for the Ising model $n=0$, the representation of dilation works for all $n$. It can be seen that $X^2=0$, so $X$ is a null vector of Cl(n+1,1), and it provides a way to represent a generic vector of Cl(n,0) in the larger conformal Clifford algebra Cl(n+1,1).

In this Appendix we want to show that the bivector $e_{\infty 0}$ generates dilations or contractions in %Cl(n+1,1), or at least in 
the subspace of Cl(n+1,1) which represents vectors in Cl(n,0). To do so, we consider an element $X$ in Cl(n+1,1) which is of the form given in Eq.~(\ref{compactification}). Because $e_{\infty 0}^2=1$, we can write
\begin{equation}\label{dilationformula}
    e^{\alpha e_{\infty 0}}=\cosh{\alpha}+e_{\infty 0}\sinh{\alpha}.
\end{equation}
Therefore, if we call
\begin{equation}
   \tilde{X_d}(x)=e^{-\alpha e_{\infty 0}/2} X(x) e^{\alpha e_{\infty 0}/2},
\end{equation}
then, by expanding $\tilde{X}_d$ using Eq.~(\ref{dilationformula}), we obtain 
%as
%\begin{equation}
%\left[\cosh{\frac{\alpha}{2}}-e_{\infty 0}\sinh{\frac{\alpha}{2}}\right] \left[x+e_0-\frac{x^2}{2}e_{\infty}\right] \left[\cosh{\frac{\alpha}{2}}+e_{\infty 0}\sinh{\frac{\alpha}{2}}\right],
%\end{equation}
%we can write
\begin{eqnarray}\label{conformaldilation}
    \tilde{X_d}(x) %& = & \left[\cosh{\frac{\alpha}{2}}-e_{\infty 0}\sinh{\frac{\alpha}{2}}\right] \left[x+e_0-\frac{x^2}{2}e_{\infty}\right] \left[\cosh{\frac{\alpha}{2}}+e_{\infty 0}\sinh{\frac{\alpha}{2}}\right] \\ \nonumber
    & = &  x \left[\cosh^2 {\frac{\alpha}{2}}-e_{\infty 0}^2\sinh^2{\frac{\alpha}{2}}\right] \\ \nonumber
     %& + & \left[e_0\cosh^2 {\frac{\alpha}{2}}+[e_0,e_{\infty 0}]\cosh {\frac{\alpha}{2}}\sin{\frac{\alpha}{2}}-e_{\infty 0}e_0 e_{\infty 0}\sinh^2 {\frac{\alpha}{2}}\right] \\ \nonumber
     %& - & \frac{x^2}{2}\left[e_{\infty}\cosh^2 {\frac{\alpha}{2}}+[e_{\infty},e_{\infty 0}]\cosh {\frac{\alpha}{2}}\sin{\frac{\alpha}{2}}-e_{\infty 0}e_\infty e_{\infty 0}\sinh^2 {\frac{\alpha}{2}}\right] \\ \nonumber
    & = & x + e_0\left[\cosh^2 {\frac{\alpha}{2}}+\sinh^2 {\frac{\alpha}{2}}-2\sinh {\frac{\alpha}{2}}\cosh {\frac{\alpha}{2}}\right] \\ \nonumber
    & - & \frac{x^2}{2} e_{\infty}\left[\cosh^2 {\frac{\alpha}{2}}+\sinh^2 {\frac{\alpha}{2}}+2\sinh {\frac{\alpha}{2}}\cosh {\frac{\alpha}{2}}\right] \\ \nonumber
    %& = & x + e_0\left[\cosh \alpha-\sinh{\alpha}\right]-\frac{x^2}{2}e_{\infty} \left[\cosh \alpha+\sinh \alpha\right] \\ \nonumber
    %& = & x +e_0 e^{-\alpha} - \frac{x^2}{2}e^{\alpha} \\ \nonumber
    & = &  e^{-\alpha}\left[e^{\alpha}x+e_0-\frac{\left(e^{\alpha}x\right)^2}{2}e_{\infty}\right]
    = e^{-\alpha} X(e^{\alpha}x),
\end{eqnarray}
where we have denoted by $[X,Y]$ the commutator of $X$ and $Y$, or $[X,Y]=XY-YX$.
Eq.~(\ref{conformaldilation}) means that the transformation which maps $X$ into $\tilde{X_d}$ in Cl(n+1,1) is equivalent to the one which maps $x$ onto $\tilde{x}=e^{\alpha}x$ in Cl(n,0), which is a dilation.

\end{document}